\documentclass[12pt]{iopart}

\usepackage{color}
\usepackage{graphicx}
\usepackage{amssymb}
\usepackage{amsmath}

\begin{document}

\title{Suppression of the radiation squeezing in interacting quantum Hall edge channels.}

\author{G Rebora$^{1,2}$, D Ferraro$^{1,2}$ and M Sassetti$^{1,2}$}

\address{$^{1}$Dipartimento di Fisica, Universit\`a di Genova, Via Dodecaneso 33, 16146, Genova, Italy}
\address{$^{2}$SPIN-CNR, Via Dodecaneso 33, 16146 Genova, Italy}

\begin{abstract}
We study the quantum fluctuations of the two quadratures of the emitted electromagnetic radiation generated by a quantum Hall device in a quantum point contact geometry. In particular, we focus our attention on the role played by the unavoidable electron-electron interactions between the two edge channels at filling factor two. We investigate quantum features of the emitted microwave radiation, such as  squeezing, by studying the current fluctuations at finite frequency, accessible through a two-filters set-up placed just after the quantum point contact. We compare two different drives, respectively a cosine and a train of Lorentzian pulses, used for the injection of the excitations into the system. In both cases quantum features are reduced due to the interactions, however the Lorentzian drive is still characterized by a robust squeezing effect which can have important application on quantum information.
\end{abstract}

\section{Introduction}
 
During the last decades, significant theoretical and experimental efforts have been directed towards the control of individual electronic degrees of freedom in mesoscopic devices. These have led to the development of electron quantum optics (EQO)~\cite{Grenier11, Bocquillon14, Roussel17, Ferraro17b, Glattli17, Bauerle18} which aims at transposing conventional quantum optics concepts to describe the manipulation and measurement of electrons propagating in quantum channels. To this purpose, a fundamental building block is represented by the realization of single electron sources, such as driven quantum dots~\cite{Feve07, Mahe10} or properly designed Lorentzian voltage pulses in time, also known as Levitons,~\cite{Dubois13b, Jullien14}, which allow to inject a controlled train of electronic wave-packets in quantum Hall (QH) edge channels where the ballistic propagation mimic the behaviour of photons in optics waveguides. A major achievement in this context was the realization of the electronic Hanbury-Brown-Twiss~\cite{Hanbury56} interferometer, which allows to access the granular nature of the particles, and the Hong-Ou-Mandel~\cite{Hong87} experiment, able to provide information about the statistical properties of the colliding particles, in the EQO context~\cite{Bocquillon12, Bocquillon13, Wahl14, Ferraro14, Freulon15, Marguerite16}.\\
Electrons strongly differ from photons due to their statistics and their charged nature. Indeed, in quantum channels, they are subjected to screened Coulomb interactions. This affects the dynamics of the excitations and dramatically emerges in QH edge channels such as those at filling factor $\nu=2$ leading to charge fractionalization and energy exchange between channels~\cite{Levkivskyi08, Altimiras09, Lunde10, Levkivskyi12}. The role of inter-edge coupling has been deeply investigated in order to understand the experimental observations and elaborate theoretical predictions of current fluctuations in an HOM set-up~\cite{Bocquillon13, Wahl14, Marguerite16, Rebora20}. Moreover, the interacting mechanisms lead to electronic decoherence and relaxation effects whose understanding can be necessary to properly characterize the role of single-electron excitation as carriers of quantum information~\cite{Gasse13, Rodriguez20, Rebora21}. \\
The study of current fluctuations (noise) at finite frequency, carried out in presence of normal tunnel junctions ~\cite{Gasse13, Forgues15, Forgues16} and Josephson junctions~\cite{Westig17}, has underlined the deep connection between this quantity and the fluctuations of the microwave radiation emitted by driven mesoscopic devices. This radiation manifests quantum features such as squeezing and entanglement of photons in the frequency domain, which can be properly engineered by controlling the form of the drive. In particular it has been shown that very narrow voltage pulses can maximize the single- and two-photon squeezing of the radiation emitted in the microwave range~\cite{Mendes15, Ferraro18}. A deep knowledge of such non-classical features of the emitted light is essential in the development of quantum information protocols \cite{Braunstein05}. A key ingredient in order to obtain squeezed radiation is the presence of non-linearities which allows to beat vacuum fluctuations along one quadrature. 
In this direction, electron-electron interaction naturally leads to such non-linearities in the current-voltage characteristics in a Quantum Point Contact (QPC) geometry \cite{Sassetti96, Ferraro10, Dolcetto16, Ronetti17}. However, a more detailed analysis of the effect of inter-edge coupling on the quantum properties of the emitted radiation is still missing.\\
In this paper, we consider a QH bar at filling factor $\nu=2$ in a QPC geometry. Here, the two edge channels interact through a short-range capacitive coupling and the injection of excitations in the system is made by means of time-dependent voltage pulses. We theoretically investigate the dynamical response of the outgoing current fluctuations, which are associated to the photo-assisted finite frequency noise, in the presence of Coulomb interactions. Despite the photo-assisted shot noise~\cite{Kapfer19} and the current correlations~\cite{Bartolomei20} in the strongly interacting fractional QH regime have been recently experimentally investigated, a detailed analysis of both these effects, in presence of two interacting edge channels, is still lacking but it is necessary because the dynamics of excitations is strongly affected by such phenomena which cannot be avoided in realistic experimental conditions. For this reason, a proper theoretical framework in the interacting regime is needed in order to pave the way towards new experiments devoted to the improvement of the quantum behaviour of the associated emitted electromagnetic field. Our work intends to fill this gap by focusing on the consequences of electron-electron interactions on the quantum properties of the emitted electromagnetic field, such as squeezing. In particular, we will show how the quantum fluctuations are enhanced by the presence of Coulomb interactions for both a cosine drive and a train of Lorentzian pulses in time. Despite having a reduction of the squeezing effect in both cases, the Lorentzian drive still leads to relevant squeezing even in presence of interaction and this is a signature of a better robustness of this drive with respect to the cosine one. Our analysis also contributes in the development of EQO, whose main purpose is to revise conventional optics experiments, by further deepening the study of unavoidable interacting effects.\\
The paper is organized as follows. In Section \ref{sec:model} we describe the model for two interacting QH edge channels in terms of the edge-magnetoplasmon scattering matrix formalism. In Section \ref{noise} we calculate the photo-assisted noise at finite frequency in presence of interacting edge channels in a QH bar at filling factor $\nu=2$. We also provide a possible experimental set-up able to access the current fluctuations. In Section \ref{Quadratures} we discuss how the current fluctuations, previously calculated, are related to the quadratures of the emitted electromagnetic fields. The effects of Coulomb interaction on the squeezed light are then studied in Section \ref{results} where a particular attention is devoted to the possibility of access optimal squeezing regimes in the two filters set-up. Finally, Section \ref{conclusions} is devoted to the conclusions. 
 

\section{\label{sec:model} Model}
\begin{figure}[ht]
\centering
\includegraphics[scale=0.4]{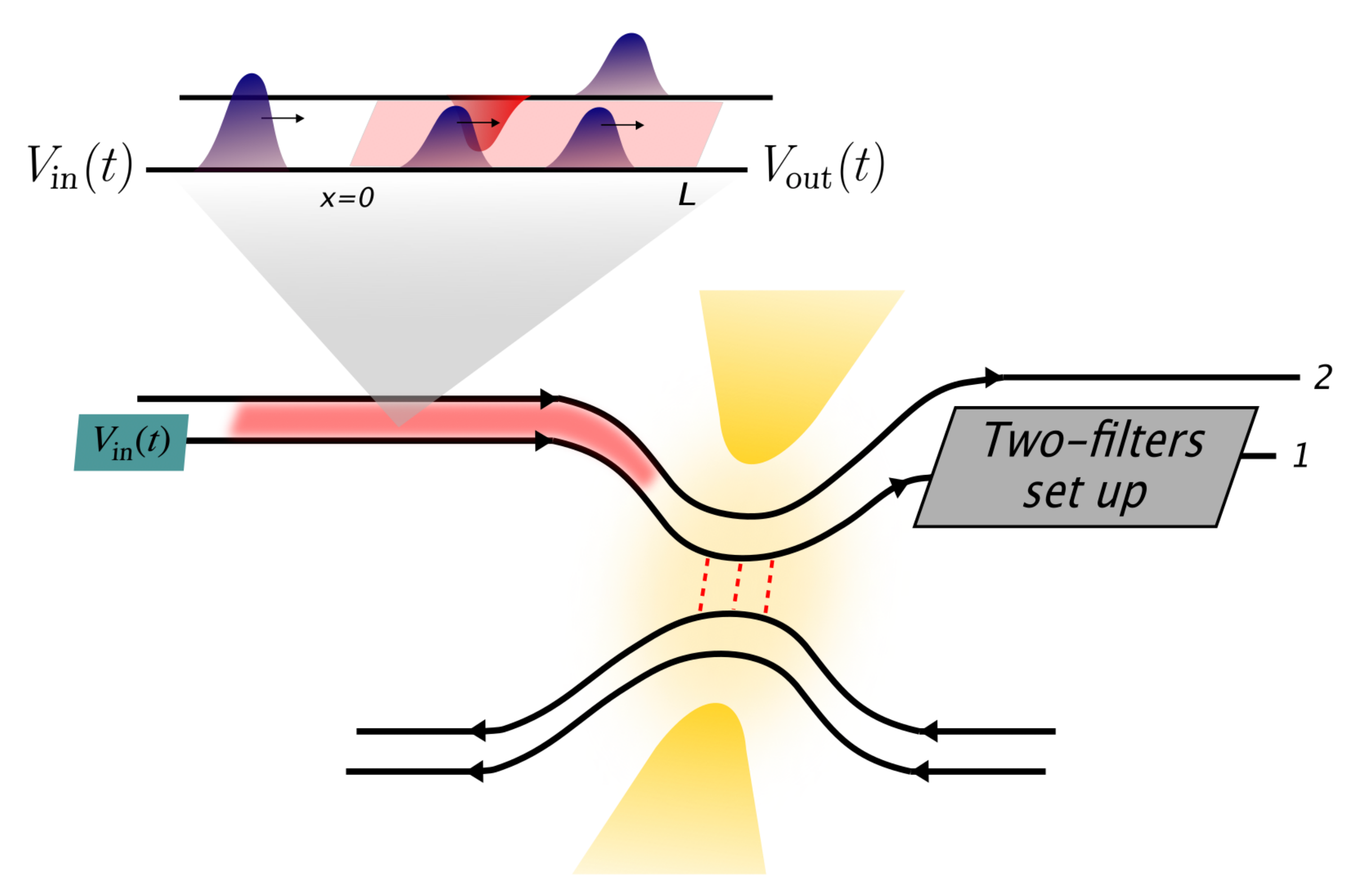}
\caption{(Color on-line) Schematic view of a quantum Hall bar at filling factor $\nu=2$ in the QPC geometry. Here, the Hall channels are capacitively coupled through a short range interaction (red) which lead to a separation of the electronic excitation, injected through the voltage source $V_{\mathrm{in}}(t)$ in the inner channel, into a fast charge and a slow neutral modes propagating with velocities $v_\rho$ and $v_\sigma$ respectively. The outgoing voltage, indicated as $V_{\mathrm{out}}(t)$, contains a DC and AC parts as in Equation~\eqref{Vout}. By means of gates (yellow), which create the QPC, the current is partitioned and reaches a two-filters set-up placed just outside the QPC region. Notice that the region of the QPC is yellow to indicate the fact that here the electron-electron interaction is negligible.} 
\label{Fig1}
\end{figure}

We consider the edge states of the upper part of a QH bar at filling factor $\nu=2$ (see Figure~\ref{Fig1}). Here, the two copropagating channels interact via a screened (ideally $\delta$-like) Coulomb repulsion~\cite{ Wahl14, Ferraro14, Levkivskyi08, Rebora20, Sukhorukov07, Degiovanni10, Ferraro17, Acciai18} over a region of finite length $L$. Following Wen's hydrodynamical description~\cite{Wen95} the behaviour of these edge states is described by the Hamiltonian density ($\hbar=1$)
\begin{equation}
    \mathcal{H}=\frac{1}{4 \pi} \sum_{p,s=1,2}\mathcal{V}_{ps}\left( \partial_x\phi_{p}(x)\right)\left( \partial_x\phi_{s}(x)\right)\, \label{eq:0}
\end{equation}
with 
\begin{equation}
  \mathcal{V}=\begin{pmatrix}
    v_{1}  & u \\
    u & v_{2}
    \end{pmatrix}.
    \label{eq:v}
\end{equation}
Here, we indicate with $1$ ($2$) the inner (outer) channel (see Figure \ref{Fig1}) and $\phi_{p}$ are chiral bosonic fields satisfying the commutation relations
\begin{equation}
[\phi_p(x),\phi_s(y)]=i \pi \,\mathrm{sign}(x-y) \delta_{ps}.
\end{equation} 
They are related to the particle density operator by~\cite{Miranda03, Ziani12, Ziani15}
\begin{equation}
\rho_{p}(x)=\frac{1}{2\pi}\partial_x \phi_{p}(x),
\end{equation} 
while the fermionic field $\psi_{p}(x)$ can be written as a coherent state of $\phi_{p}(x)$ via the conventional bosonized expression
\begin{equation}
    \psi_{p}(x) = \frac{\mathcal{F}_{p}}{\sqrt{2\pi a}}e^{-i\phi_{p}(x)}
\end{equation}
where $a$ is a short distance cut-off and $\mathcal{F}_{p}$ are the Klein factors~\cite{Wen95, Miranda03}. In Equation~(\ref{eq:v}) the parameters $v_{p}$ are the bare propagation velocities of the two edge modes and $u$ is the intensity of the capacitive coupling between them. In the following, for notational convenience and without loss of generality, we will consider $v_{2}=v$ and $v_{1}=\alpha v$ with $\alpha\geq 1$.\\
It is possible to diagonalize $\mathcal{H}$ by means of a rotation in the bosonic fields space with angle $\theta$ satisfying
\begin{equation}
\theta=\frac{1}{2}\arctan\left[\frac{2u}{(\alpha-1)v}\right]. 
\end{equation} 
Notice that at $\theta=0$ the two channels are independent ($u=0$), while $\theta=\pi/4$ represents the so called "strongly interacting" regime~\cite{ Wahl14, Ferraro14, Levkivskyi08, Degiovanni10, Ferraro17, Acciai18, Kovrizhin12, Slobodeniuk16}. This condition can be achieved for $\alpha=1$ (edge modes with identical velocities) and in principle also for $u\rightarrow +\infty$. However, the stability of the model (Hamiltonian density bounded from below) constraints the maximal admissible values of $u$ \cite{Ferraro17}. After diagonalization, the Hamiltonian density reads
\begin{equation}
    \mathcal{H}=\sum_{\beta=\rho,\sigma}\frac{v_\beta}{4\pi}(\partial_x\phi_\beta(x))^2\,,
\end{equation}
where the two eigenmodes are a fast charged field $\phi_\rho$ and a slow neutral one $ \phi_\sigma$ propagating with velocities 
\begin{equation}
    v_{\rho/\sigma}=v \left[\left(\frac{\alpha+1}{2}\right)\pm\frac{1}{\cos(2\theta)}\left(\frac{\alpha-1}{2}\right)\right]
\end{equation}
respectively.\\
The dynamics of the edge channels can be solved within the edge-magnetoplasmon scattering formalism~\cite{Ferraro14, Rebora21, Degiovanni10, Ferraro17, Sukhorukov15} where, once expressed in Fourier transform with respect to the time, the bosonic fields outgoing from the interacting region and before the QPC are related to the incoming ones through the edge-magnetoplasmon scattering matrix relation
\begin{equation}
    \begin{pmatrix}
    \phi_{1}(L,\omega) \\
    \phi_{2}(L,\omega)
    \end{pmatrix}=\mathcal{S}(L,\omega)\begin{pmatrix}
    \phi_{1}(0,\omega) \\
    \phi_{2}(0,\omega)
    \end{pmatrix}
    \label{eq:phi-out}
\end{equation}
with scattering matrix
\begin{equation}
\mathcal{S}(L,\omega)=
\left( 
\begin{matrix}
\cos^{2}\theta e^{i \omega  \tau_{\rho}}+\sin^{2}\theta e^{i \omega \tau_{\sigma}} &  \sin \theta \cos \theta \left(e^{i \omega  \tau_{\rho}}-e^{i \omega  \tau_{\sigma}} \right)\\
\sin \theta \cos \theta \left(e^{i \omega \tau_{\rho}}-e^{i \omega  \tau_{\sigma}} \right)&  \sin^{2}\theta e^{i \omega \tau_{\rho}}+\cos^{2}\theta e^{i \omega \tau_{\sigma}}\\
\end{matrix}
\right).
\label{S_matrix}
\end{equation}
In the above Equation we have introduced the times of flight associated with fast and slow modes 
\begin{equation}
\tau_{\rho/\sigma}=\frac{L}{v_{\rho/\sigma}}.
\end{equation}

We want now to consider the effects a time-dependent periodic voltage $V_{\mathrm{in}}(t)$ applied to the inner channel just before the interacting region (see Figure \ref{Fig1}). This can be done through the conventional minimal coupling Hamiltonian density
\begin{equation}
\mathcal{H}_{U}=-e\, \rho_{1}(x)U(x,t) \,,
\end{equation}
where $-e$ $(e>0)$ is the electron charge and
\begin{equation}
U(x,t)=\Theta(-x)V_{\mathrm{in}}(t)
\end{equation}
where the Heaviside step function $\Theta(-x)$ indicates the region where this potential is applied~\cite{Rech17, Vannucci17, Ronetti18, Ronetti19}. Here, we have considered the decomposition in terms of DC and AC part
\begin{equation}
V_{\mathrm{in}}(t)=V_{DC}+V_{AC}(t)
\end{equation}
with 
\begin{equation}
\frac{1}{\mathcal{T}}\int^{+\frac{\mathcal{T}}{2}}_{-\frac{\mathcal{T}}{2}} dt V_{AC}(t)=0
\label{AC_zero}
\end{equation}
being $\mathcal{T}$ the period of the drive. In the following we will focus on both a cosine drive with 
\begin{equation}
V^{(c)}_{AC}(t) = -\tilde{V} \cos \left( \frac{2 \pi t}{\mathcal{T}} \right)
\label{cosine}
\end{equation}
or a periodic train of Lorentzian pulses with width $w=\eta \mathcal{T}$
\begin{equation}
V^{(l)}_{AC}(t) = \frac{\tilde{V}}{\pi} \sum_{l=-\infty}^{+\infty} \frac{\eta}{\eta^2 + \left( \frac{t}{\mathcal{T}}-l \right)^2}-\tilde{V}.
\label{Lorentzian}
\end{equation}
In both cases $\tilde{V}$ is the amplitude of the AC part of the drive. By solving the equations of motion for the complete Hamiltonian density $\mathcal{H}+\mathcal{H}_{U}$~\cite{Safi99, Grenier13} one obtains the voltage outgoing from the interacting region which is subject to neutral-charge modes separation in the following way~\cite{Freulon15, Levkivskyi08, Safi99}
\begin{equation}
V_{\mathrm{out}}(t)= V_{DC}+\cos^{2}{\theta} V_{AC}(t-\tau_{\rho})+\sin^{2}{\theta}V_{AC}(t-\tau_{\sigma}).
\label{Vout}
\end{equation}
Figure~\ref{Fig1} shows that the excitations emitted by the periodic voltage source, flowing along the edge, fractionalize through the interacting region and then are partitioned at a QPC where only inner channels are involved in the tunneling. This configuration can be implemented by properly tuning the QPC gate's voltages~\cite{Bocquillon14, Wahl14, Marguerite16}. Moreover, the partitioning region of the QPC is not included in the interacting region as long as both the interedge interaction and the tunneling are local (see for example Supplementary Material of Ref.~\cite{Wahl14}). This description is physically motivated by the fact that both the contacts used to apply the voltage and the gates that realize the QPC locally enhance the screening of the interaction and so the colliding fermions are locally free at this location. The free fermionic fields, outgoing from the QPC (denoted with \textit{after}), are related to the incoming ones (\textit{before}) through
\begin{equation}
\begin{pmatrix}
\psi_R \\ \psi_L
\end{pmatrix}_{(\textit{after})}
=
\begin{pmatrix}
\sqrt{\mathcal{T}} & i\sqrt{\mathcal{R}} \\
i\sqrt{\mathcal{R}} & \sqrt{\mathcal{T}}
\end{pmatrix}
\begin{pmatrix}
\psi_R \\ \psi_L
\end{pmatrix}_{(\textit{before})}
\label{qpc}
\end{equation}
where $\psi_{R,L}$ are, respectively, the right- and left-mover fermionic fields, flowing along the inner channels (1) and the coefficient $\mathcal{R}$ and $\mathcal{T}=1-\mathcal{R}$ are the reflection and transmission probabilities of the QPC. According to these considerations, to the chirality and to the locality of the coupling, we can consider the interacting region extending from just after the injection point to just before the QPC~\cite{Ferraro14, Degiovanni10, Ferraro17}.\\
The inner channel of the upper edge enter into a two-filters set-up which is placed immediately after the QPC in such a way that the interactions between the outgoing channels are there negligible. In the following Sections we will investigate the current which is subjected to fluctuations (noise) and we will focus on the zero temperature limit. 


\section{Current and photo-assisted noise at finite frequency}\label{noise}
\begin{figure}[ht]
\centering
\includegraphics[scale=0.4]{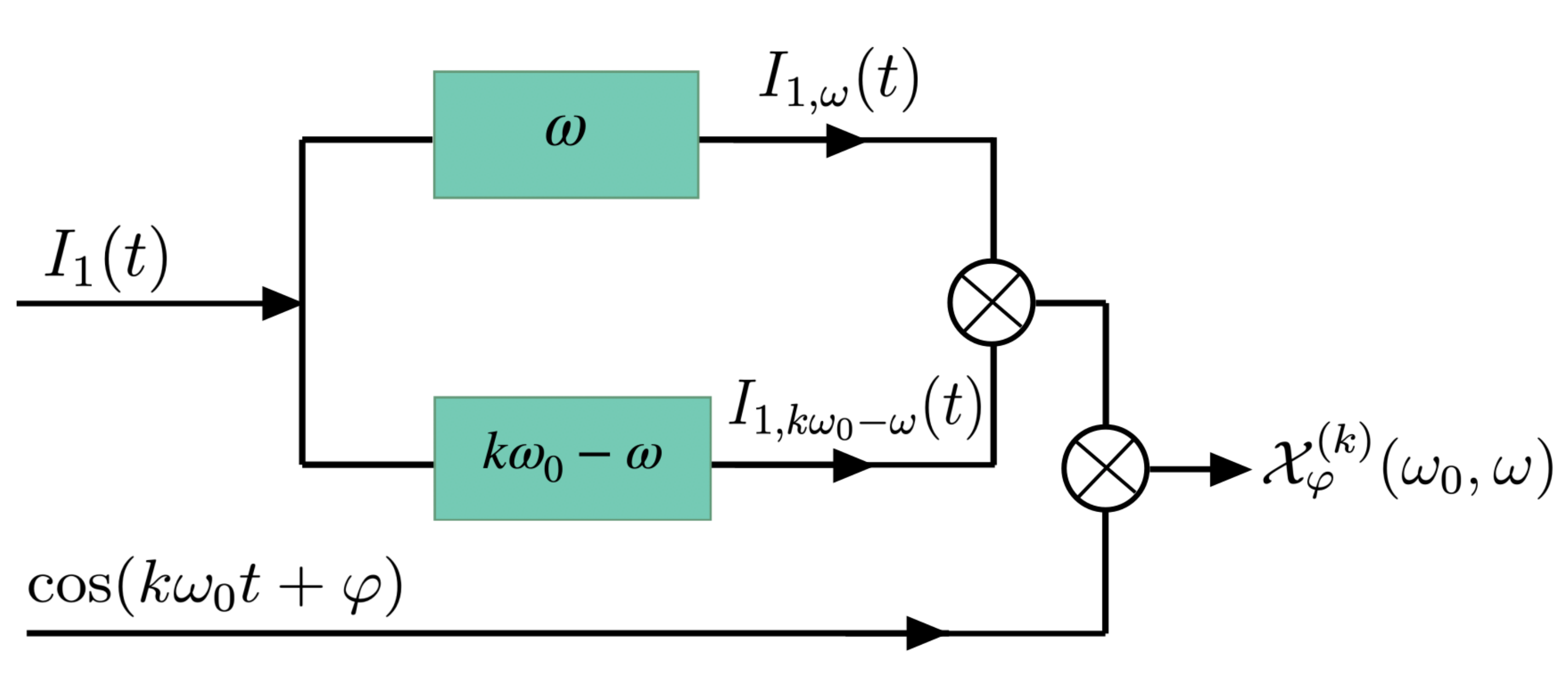}
\caption{(Color on-line) Schematic view of a set-up designed to measure the dynamical response of the noise $\mathcal{X}_{\varphi}^{(k)}$. The current of the inner channel outgoing from the QPC (associated to the operator $I_1(t)$) of the Hall bar is further split into two at the level of the detector. The resulting contributions are filtered at the two frequencies $\omega$ ($I_{1,\omega}(t)$) and $|k \omega_{0}-\omega|$ ($I_{1,k\omega_0-\omega}(t)$) respectively (green boxes). They are then multiplied among themselves and with a cosine signal ($\otimes$ symbols) with a tunable phase shift $\varphi$. The final output is then averaged over time.}
\label{Fig2}
\end{figure}  
Following what is usually investigated in experiments, we focus our attention on the evaluation of the finite frequency noise $S(\omega_1,\omega_2)$ defined as~\cite{Blanter00, Martin05, Ferraro14b}
\begin{equation}
    S(\omega_1,\omega_2)=\int_{-\infty}^{+\infty} dt \int_{-\infty}^{+\infty} dt' \,e^{i\omega_1 t} e^{i\omega_2 t'} C(t,t')
\end{equation}
where the current-current correlator is
\begin{equation}
    C(t,t')=\langle I_1(t)I_1(t')\rangle_c=\langle I_1(t)I_1(t')\rangle-\langle I_1(t)\rangle\langle I_1(t')\rangle.
\end{equation}
Here the notation $\langle ... \rangle_c$ stands for the connected zero temperature quantum mechanical correlator and the current operator of the inner channel is $I_1(t)=-ev_1:\psi_1^\dagger(t)\psi_1(t):$, where $:...:$ denotes the normal ordering with respect to the Fermi sea and the fermionic fields $\psi_1$ are evaluated at the level of the two-filters set-up after the QPC.\\ 
In order to access the stationary component of the photo-assisted noise at finite frequency~\cite{Blanter00, Martin05, Lesovik94, Schoelkopf98} (zero-th order harmonic) and the more general dynamical response of the outgoing current fluctuations~\cite{ Ferraro18, Gabelli08, Gabelli08b}, corresponding to the higher order harmonics at a frequency $k\omega_{0}$ ($k \in \mathbb N$), we consider a two-filter measurement device, placed immediately after the QPC (see Figure \ref{Fig2}). As depicted in Figure~\ref{Fig2}, the current of the inner channel (associated to the operator $I_1(t)$), incoming the set-up, is filtered at two different frequencies, respectively $\omega$ and $|k\omega_0-\omega|$. By considering one of the two filter, for example the one in $\omega$, the outgoing current becomes
\begin{equation}
    I_{1}(t)\rightarrow I_{1,\omega}(t)\approx I_1(\omega)e^{i\omega t}+I_1(-\omega)e^{-i\omega t}.
    \label{filter}
\end{equation}
The filtered contributions are then multiplied among themselves and further with a cosine signal.\\
According to the described set-up and to the symmetrized correlators of the current operators~\cite{Ferraro18, Gabelli08, Ferraro14b}, the $k$-th harmonics of the dynamical response of the noise measured at a frequency $\omega$ is given by
\begin{equation}
\mathcal{X}^{(k)}_{\varphi}(\omega_{0}, \omega)=\lim_{\mathcal{T}_{m} \rightarrow +\infty}\frac{1}{\mathcal{T}_{m}} \int^{+\frac{\mathcal{T}_{m}}{2}}_{-\frac{\mathcal{T}_{m}}{2}} dt \langle I_{1,\omega}(t) I_{1,k\omega_{0}-\omega}(t) \cos(k \omega_{0}t+\varphi) \rangle_c, 
\label{X_k_generic} 
\end{equation}
with $I_{1,\omega}(t)$ defined in Equation~\eqref{filter} and 
\begin{equation}
\omega_{0}=\frac{2 \pi}{\mathcal{T}}\, .
\end{equation}
Notice that we consider an averaging over a given measurement time $\mathcal{T}_{m}$ typically longer with respect to all the other time scales involved in the dynamics of the system. Therefore, keeping only the non zero terms in the average, the expression in Equation~\eqref{X_k_generic} reduces to
\begin{equation}
    \mathcal{X}^{(k)}_\varphi(\omega_0,\omega)=\frac{1}{2}[\langle I_1(\omega)I_1(k\omega_0 -\omega)e^{-i\varphi}\rangle_c+\langle I_1(-\omega)I_1(-k\omega_0+\omega)e^{i\varphi}\rangle_c].
\end{equation}
According to this and to the reality constraint $I_1^{\dagger}(\omega)=I_1(-\omega)$, for the current operator, one can write 
\begin{equation}
\mathcal{X}^{(k)}_{\varphi}(\omega_{0}, \omega)= \frac{1}{2} \left[\Re\left\{\mathcal{X}^{(k)}_{+, \varphi}(\omega_{0}, \omega)\right\}+\Re\left\{\mathcal{X}^{(-k)}_{+,\varphi}(\omega_{0}, -\omega)\right\}\right] \label{Xp}
\end{equation}
with
\begin{equation}
\mathcal{X}^{(k)}_{+, \varphi}(\omega_{0}, \omega)=e^{i\varphi}\langle I_1(\omega) I_1(k\omega_{0}-\omega) \rangle_{c}
\label{corr}
\end{equation}
and $\Re\left\{...\right\}$ indicating the real part. From this point we will consider only $k>0$ by using the properties of $\mathcal{X}^{(\pm k)}_{+,\varphi}(\omega_0,\pm \omega)$. The expression for the dynamical response of the noise $\mathcal{X}^{(k)}_\varphi(\omega_0,\omega)$, evaluated explicitly in~\ref{appendixA} when a periodic drive is applied to the inner edge channel (1), becomes
\begin{equation}
\mathcal{X}^{(k)}_{\varphi}(\omega_0,\omega)=\frac{\cos{\varphi}}{2}\, \Re\left\{\sum_{n=-\infty}^{+\infty} \widetilde{p}_{n}(z) \widetilde{p}^{*}_{n+k}(z) S_{0}(q+1+n\lambda)+  \sum_{n=-\infty}^{+\infty} \widetilde{p}_{n}(z) \widetilde{p}^{*}_{n-k}(z) S_{0}(q-1+n\lambda)  \right\}.
\label{Xp_extended}
\end{equation}
Notice that, in the above expression we have introduced the short-hand notations: $q=e V_{DC}/\omega$, $\lambda=\omega_{0}/\omega$ and $z=e \tilde{V}/\omega_{0}$. The quantity \begin{equation}
S_{0}(\xi)= \mathcal{A}(\omega) |\xi|
\end{equation}
represents the shot noise \cite{Blanter00, Martin05} with 
\begin{equation}
\mathcal{A}(\omega)=G \mathcal{R}\mathcal{T} \omega
\end{equation}
where $G=e^{2}/2 \pi$ is the linear conductance of the considered channel and $\mathcal{R}\mathcal{T}$ is the product of the reflection and transmission coefficients of the quantum point contact. 
 
Moreover, according to the general definition reported in Ref. \cite{Rebora20} one has
\begin{equation}
  \widetilde{p}_{l}(z)=\sum_{n} p_{l-n}(z_1)p_{n}(z_2) e^{i \omega_{0}\tau_\rho (l-n)} e^{i \omega_{0}\tau_\sigma (n)}
\end{equation}
where $z_1=z\cos^2\theta $ and $z_2=z\sin^2\theta $. Thus, the coefficients $\widetilde{p}_{l}(z)$ are completely specified once the expression of $p_{n}(z)$ is known. These are the photo-assisted amplitudes \cite{Dubois13, Crepieux04}, defined as
\begin{equation}
p_{n}(z) = \int_{-\frac{\mathcal{T}}{2}}^{+\frac{\mathcal{T}}{2}} \frac{dt}{\mathcal{T}} e^{2 i \pi  n \frac{t}{\mathcal{T}}} e^{- 2 i\pi  z  \Phi(t)}, 
\label{eq:defpofl}
\end{equation}
with 
\begin{equation}
\Phi (t) = e \int_{-\infty}^t \frac{dt'}{\mathcal{T}}\frac{V_{AC}(t')}{\tilde{V}}.
\end{equation}

It is worth to note that for $k=0$ and $\varphi=0$, the expression in Equation (\ref{Xp}) reduces to
\begin{equation}
\mathcal{X}^{(0)}_{0}(\omega_0,\omega)=\frac{1}{2} \sum_{n=-\infty}^{+\infty}|\widetilde{p}_{n}(z)|^2 \left[S_{0}(q+1+n\lambda)+ S_{0}(q-1+n\lambda)\right],
\label{S_tilde}
\end{equation}
which represents the photo-assisted noise measured at finite frequency $\omega$ \cite{Gasse13}.

In order to proceed, we need to specify the functional form of the photo-assisted amplitudes $p_{n}(z)$ for the considered voltage drives. In the case of the cosine drive in Equation (\ref{cosine}) one has
\begin{equation}
p^{(c)}_{n}(z) = J_{n} \left(-z\right).
\label{cos}
\end{equation}
with $J_{n}(x)$ the $n$-th Bessel's function of the first kind, while in the case of the Lorentzian voltage pulse in Equation (\ref{Lorentzian}) one has \cite{Grenier13, Dubois13}
\begin{equation}
p^{(l)}_{n}(z) =  z \sum_{s=0}^{+\infty} \frac{\Gamma (z+n+s)}{\Gamma(z+1-s)} \frac{(-1)^s e^{-2 \pi \eta (2s+n)}}{(n+s)! s!},
\label{lor}
\end{equation}
with $\Gamma(x)$ the Euler's Gamma function. 
\section{Current fluctuations and electromagnetic quadratures}\label{Quadratures} 
The outgoing current operator $I_1$ can be also related to the emitted electromagnetic field annihilation operator $a$ through the relation\footnote{It is worth to point out that, in an interacting one dimensional chiral channel, an analogous relation (up to a numerical prefactor) can be derived connecting the current with the edge-magnetoplasmon creation (annihilation) operator. Therefore the conclusion derived in the following can be generalized also to this bosonic operators \cite{Feve_private}.} \cite{Grimsmo16}
\begin{equation}
a(\omega)= -i \frac{I_1(\omega)}{\sqrt{2 \mathcal{A}(\omega)}}.
\end{equation}
According to this, the electromagnetic field can be written as \cite{Gasse13, Mendes15}
\begin{align}
A_{\varphi}(\omega)&=\frac{1}{\sqrt{2}}\left[e^{i\varphi}I_1(\omega) +e^{-i \varphi} I_1(-\omega)\right]\nonumber \\
&=i \sqrt{\mathcal{A}(\omega)} \left[e^{i\varphi}a(\omega)-e^{-i \varphi}a^{\dagger}(\omega) \right]
\label{A}
\end{align}
where the phase $\varphi$ between $a$ and $a^{\dagger}$ allows to span the whole phase space. By using the Robertson formulation of the Heisenberg principle for two generic operators
\begin{equation}
    \Delta A \Delta B \geq \frac{1}{2}|\langle [A,B] \rangle|
\end{equation}
the fluctuations of the quadratures of the electromagnetic field satisfy the uncertainty relation 
\begin{equation}
\Delta A_{2\varphi} \Delta A_{2\varphi+\pi} \geq \mathcal{A}(\omega),
\label{heisen}
\end{equation}
with $\Delta A_{2\varphi}=\sqrt{\langle A_{2\varphi}^{2}\rangle-\langle A_{2\varphi} \rangle^{2}}$, which naturally link the quantum fluctuations of the electromagnetic field quadrature at a given frequency $\omega$ with the current fluctuations (see Equation~\eqref{II}) and can be detected by the experimental scheme previously presented. The effect of the radiation squeezing results from the previous Equation~\eqref{heisen}: when the value of the fluctuations of one quadrature decreases below the quantum vacuum the other must increases in order to preserve the uncertainty relation. This means that if we know with a great precision $\Delta A_{2\varphi}$ we must give hope of knowing precisely the value of the other ($\Delta A_{2\varphi+\pi}$). Furthermore it can be observed, from the definition in the second line of Equation~\eqref{A}, in terms of creation and annihilation operator, that the expectation value $\langle A_{2\varphi}\rangle$ on the equilibrium state is always zero.\\
In order to bind the dynamical response of the noise, measurable through the two-filters set-up, to the quadrature fluctuations we use the first line of Equation~\eqref{A} to calculate
\begin{equation}
\begin{split}
    \langle A^2_{2\varphi}(\omega)\rangle&=\frac{1}{2}\langle\bigg[I_1(\omega)e^{i\varphi}+I_1(-\omega)e^{-i\varphi}\bigg]\bigg[I_1(\omega)e^{i\varphi}+I_1(-\omega)e^{-i\varphi}\bigg]\rangle=\\
    &=\frac{1}{2}\langle\bigg[I_1(\omega)I_1(-\omega)+I_1(-\omega)I_1(\omega)+e^{2i\varphi}I_1(\omega)I_1(\omega)+e^{-2i\varphi}I_1(-\omega)I_1(-\omega)\bigg]\rangle .
    \label{II}
\end{split}
\end{equation}
We immediately notice that the first two terms correspond to the photo-assisted noise $\mathcal{X}_0^{(0)}(\omega_0,\omega)$ at finite frequency while the other two are equal, up to a proportional term in the phase, to the expression in Equation~\eqref{Xp} under the general constraint $k\omega_0=2\omega$. In the following we will focus of the case $k=1$ where a greater squeezing can be achieved \cite{Gasse13, Ferraro18}. Furthermore, since we are dealing with a low energy theory for the bosonic modes and the high-frequency noise generated by the stream of single electrons gets suppressed with increasing frequency~\cite{Parmentier12, Moskalets13}, we have to consider as an upper limit for the possible frequencies the energy gap $E_{g}$ of the QH state. Therefore, in order to efficiently investigate the squeezing of the emitted microwave radiation using fluctuations of an electric current we cannot exceed few GHz in the drive frequency.\\
Following Equation~\eqref{II}, the quadratures of the emitted electromagnetic field can be written in a more general way as
\begin{equation}
  \langle A^2_{2\varphi} (\omega)\rangle =\mathcal{X}_{0}^{(0)}(2\omega,\omega)+\mathcal{X}_{2\varphi}^{(1)}(2\omega,\omega)
    \label{phi}
\end{equation}
where $\varphi$ is the phase of the cosine signal of the measurement set-up in Figure~\ref{Fig2}.
In the following we will focus on the two orthogonal quadratures $\langle A^2_{0} \rangle$ and $\langle A^2_{\pi} \rangle$. Furthermore this configuration leads to the maximal squeezing in the non interacting case \cite{Gasse13}.
\section{Results}\label{results}
\begin{figure}[ht]
\centering
\includegraphics[scale=0.37]{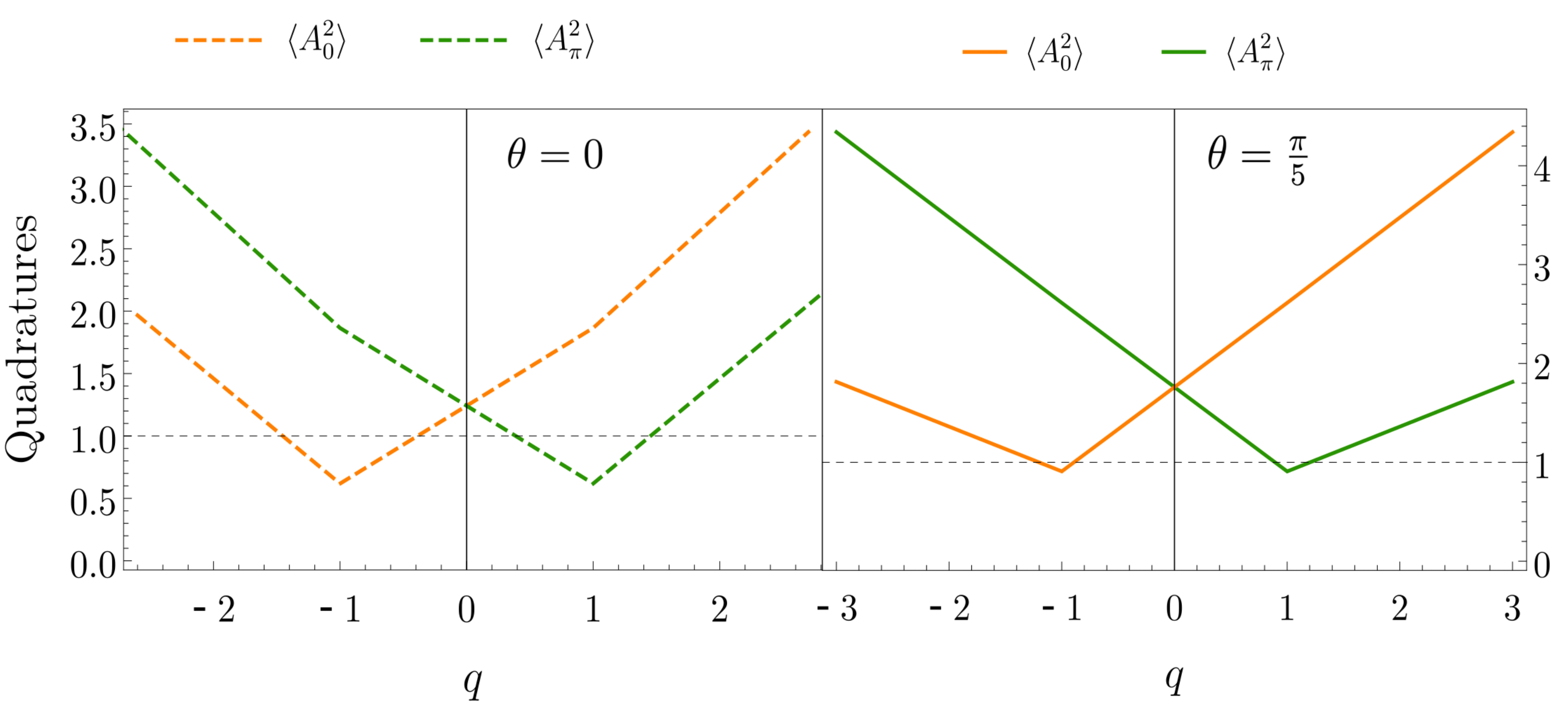}
\caption{(Color on-line) Quadratures of the emitted electromagnetic field in units of $\mathcal{A}(\omega)$ as a function of the number of injected electrons $q=eV_{\mathrm{DC}}/\hbar\omega$ for a cosine drive of period $\mathcal{T}$. Both panels represent the two orthogonal quadratures $\langle A_0^2\rangle$ (orange) and $\langle A_\pi^2\rangle$ (green): the non-interacting case ($\theta=0$) with $z=e\tilde{V}/\hbar\omega_0=0.706$ is described in the left panel (dashed curves). The right one (full curves) describes the quadratures in presence of interactions ($\theta=\pi/5$) and for $z=1.32$. The dashed gray horizontal line indicates the vacuum fluctuations. The choice of the parameters $z$ in the two cases is motivated by a numerical minimization of the noise. Other parameters are: $L=2.5\,\mu\mathrm{m}$, $v=2.8\times 10^4\, \mathrm{m/s}$, $\alpha=2.1$ and $\lambda=2$.
} 
\label{Fig3}
\end{figure}

\begin{figure}[ht]
\centering
\includegraphics[scale=0.35]{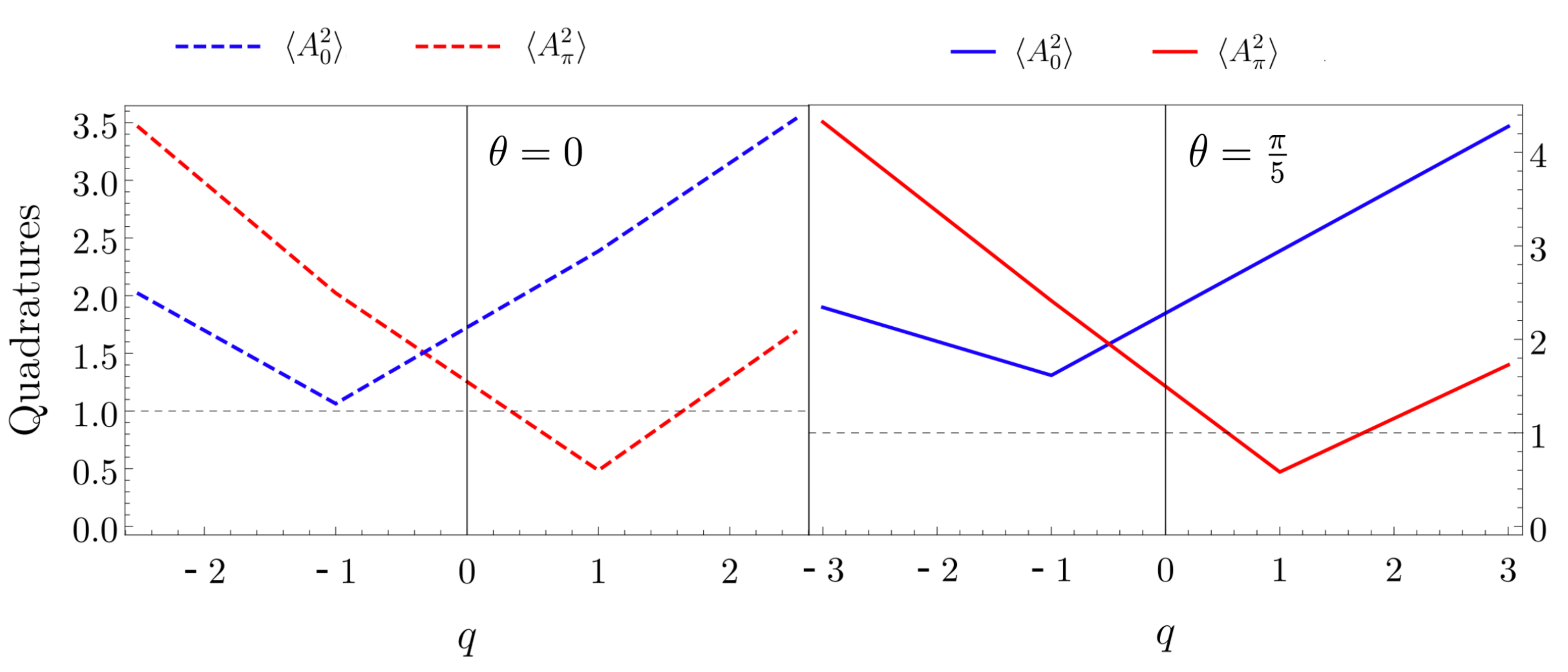}
\caption{(Color on-line) Quadratures of the emitted electromagnetic field in units of $\mathcal{A}(\omega)$ as a function of the number of injected electrons $q=eV_{\mathrm{DC}}/\hbar\omega$. The two panels describe the orthogonal quadratures $\langle A_0^2\rangle$ (blue) and $\langle A_\pi^2\rangle$ (red) for a Lorentzian voltage drive with $\eta=0.1$. The left panel (dashed curves) represents the quadratures in absence of interactions and with $z=e\tilde{V}/\hbar\omega_0=0.856$. The right one (full curves) describes the quadratures in presence of interactions ($\theta=\pi/5$) and for $z=1.21$. The dashed gray horizontal line indicates the vacuum fluctuations. The choice of the parameters $z$ in the two cases is motivated by a numerical minimization of the noise. Other parameters are: $L=2.5\,\mu\mathrm{m}$, $v=2.8\times 10^4\, \mathrm{m/s}$, $\alpha=2.1$ and $\lambda=2$.
} 
\label{Fig4}
\end{figure}

\begin{figure}[ht]
\centering
\includegraphics[scale=0.7]{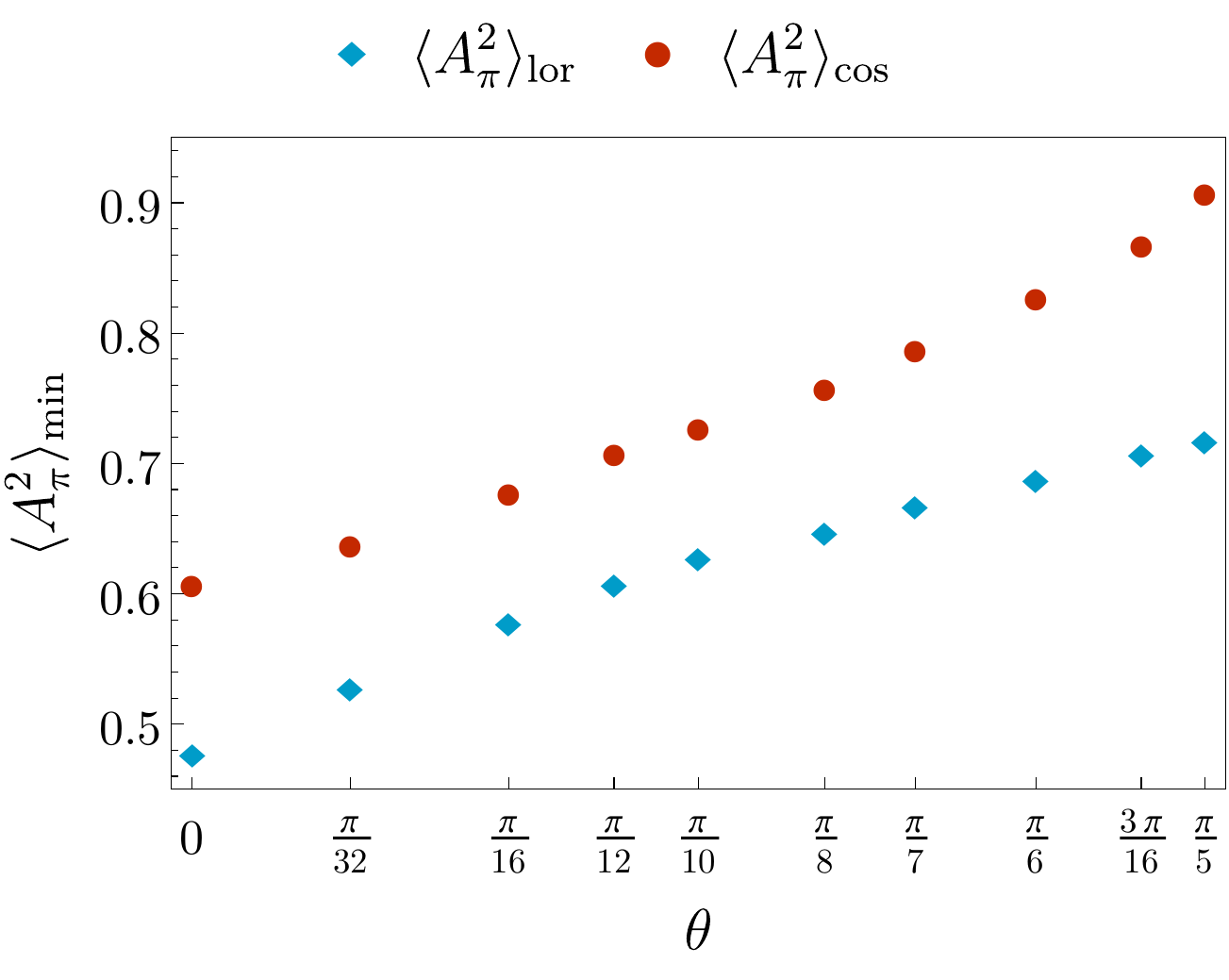}
\caption{(Color on-line) Evolution of the minimum of the $\langle A_{\pi}^2 \rangle$ quadrature, in units of $\mathcal{A}(\omega)$, as a function of the interaction parameter $\theta$ for the two different voltage drives evaluated at $q=1$. The light blue diamonds describe the behavior of $\langle A_{\pi}^2 \rangle$ when a Lorentzian drive is applied while the red circles are referred to a cosine drive. In presence of interactions, the minimum of the quadrature increases for both drives. However, a train of Lorentzian pulses is still a better candidate in order to maximize the squeezing. Notice that, for all points, the $z$ values are chosen in order to numerically minimize the noise. Other parameters are: $L=2.5\,\mu\mathrm{m}$, $v=2.8\times 10^4\, \mathrm{m/s}$, $\alpha=2.1$ and $\lambda=2$.} 
\label{Fig5}
\end{figure}

In this Section we will show how the effects of Coulomb interactions affect the squeezing of the emitted electromagnetic field through the study of the quadratures $\langle A^2_{2\varphi}\rangle$ and the related fluctuations $\Delta A_{2\varphi}$. In this context, the squeezing property offers the possibility to beat the vacuum fluctuations along one quadrature at the expense of the other in order to preserve the inequality in Equation~\eqref{heisen}. Due to these considerations the squeezing of the electromagnetic field is achieved for $\langle A^2_{2\varphi}\rangle/\mathcal{A}(\omega)<1$.\\
In Figure~\ref{Fig3} and Figure~\ref{Fig4}, we present the two orthogonal quadratures $\langle A^2_{0}\rangle$ and $\langle A^2_{\pi}\rangle$ of the emitted electromagnetic field for the cosine drive, considered in Equation~\eqref{cosine}, and for the Lorentzian one, in Equation~\eqref{Lorentzian}. The two panels of both figures allow us to compare the behaviour of the quadratures in the non-interacting case (dashed curves in the left panels) with the interacting one (full curves in the right panels). It can be immediately observed that, for both the considered AC voltage drives, the effects of interactions lead to a reduction of the value of the minima of the two orthogonal quadratures due to the neutral-charge modes separation and the consequently rising of the quantum noise outgoing from the QPC.\\
In the cosine case for a non-zero $\theta$ (right panel of Figure (\ref{Fig3})), it is evident the almost total suppression of the squeezing because the minimum of the quadrature $\langle {A}^2_0\rangle$ ($\langle A_\pi^2\rangle$) is only slightly below the quantum vacuum (horizontal gray dashed line) while the other stays well above. Concerning Figure~\ref{Fig3}, the symmetry between the two quadratures for ($q\rightarrow -q$), observed in the non-interacting case \cite{Ferraro18}, is still preserved in the interacting one according to the general properties of the photo-assisted amplitudes~\cite{Dubois13, Ronetti19b}.\\
When we turn our attention to the Lorentzian case, in Figure~\ref{Fig4}, the situation appears quite different. First of all, the mirror symmetry encountered for the cosine drive is absent here~\cite{Dubois13, Ronetti19b}. Then, the interactions still increase the values of the quadratures' minima but $\langle A^2_{\pi}\rangle$, for $q=1$, continues to stay well below the quantum vacuum limit. It is a clear sign that even in the presence of electron-electron interactions the squeezing effect is not suppressed when the injection is done with a Lorentzian drive which appears more robust and remains better with respect to the cosine one, in order to maximize the squeezing (minimizing the noise). Furthermore, the correspondent conjugate quadrature $\langle{A_0^2}\rangle$ is well above this limit as expected from a generic squeezed state where the fluctuations along one quadrature are suppressed while the others are enhanced in order to obey the Heisenberg's uncertainty relation. The presented results, for all panels in Figures~\ref{Fig3} and~\ref{Fig4}, have been obtained by numerically minimizing the noise as a function of $z$ which is the amplitude of the AC drive. This procedure is done on the one hand because the interactions deform the pulse of the AC drive and these effects have to be properly compensated and one the other hand in order to find the amplitude of the drive leading to the optimal squeezing. It is worth noting that by considering the optimal value of $z$ for the cosine drive, the squeezing associated to the Lorentzian drive is still slightly better. This comparison has been done in~\ref{app2}. Hence it follows that we have used different values of $z$ for the considered drive in the non-interacting case with respect to the interacting one.\\
Figure~\ref{Fig5} shows the behaviour of the minimum (i.e. for $q=1$) of $\langle A_\pi^2\rangle$ as a function of the interaction coupling strength $\theta$ for the two drives considered previously. We observe that, by increasing the interactions, the minimum of the quadrature increases its value for both cosine drive (light blue diamonds) and Lorentzian one (red circles). However, the squeezing effect is always better if we consider the injection of Levitons with respect to a cosine drive and it can be seen for all the considered interaction strengths. It is worth to notice that for all points we have chosen a value for $z$ which minimize the noise.\\
Until now, recalling Figure~\ref{Fig2}, we have considered a fixed value of $\varphi$ for both quadratures respectively $\langle A_{0}^2 \rangle$ ($\varphi=0$) and $\langle A_{\pi}^2 \rangle$ ($\varphi=\pi/2$). We remind that this is the phase of the cosine signal which is multiplied with the filtered current in the two-filters set-up. From this point of view, we can consider the phase as an experimentally tunable parameter and we might ask what happens to the quantum fluctuations $\Delta A_{2\varphi}$ when $\varphi$ changes.\\
\begin{figure}[ht]
\centering
\includegraphics[scale=0.38]{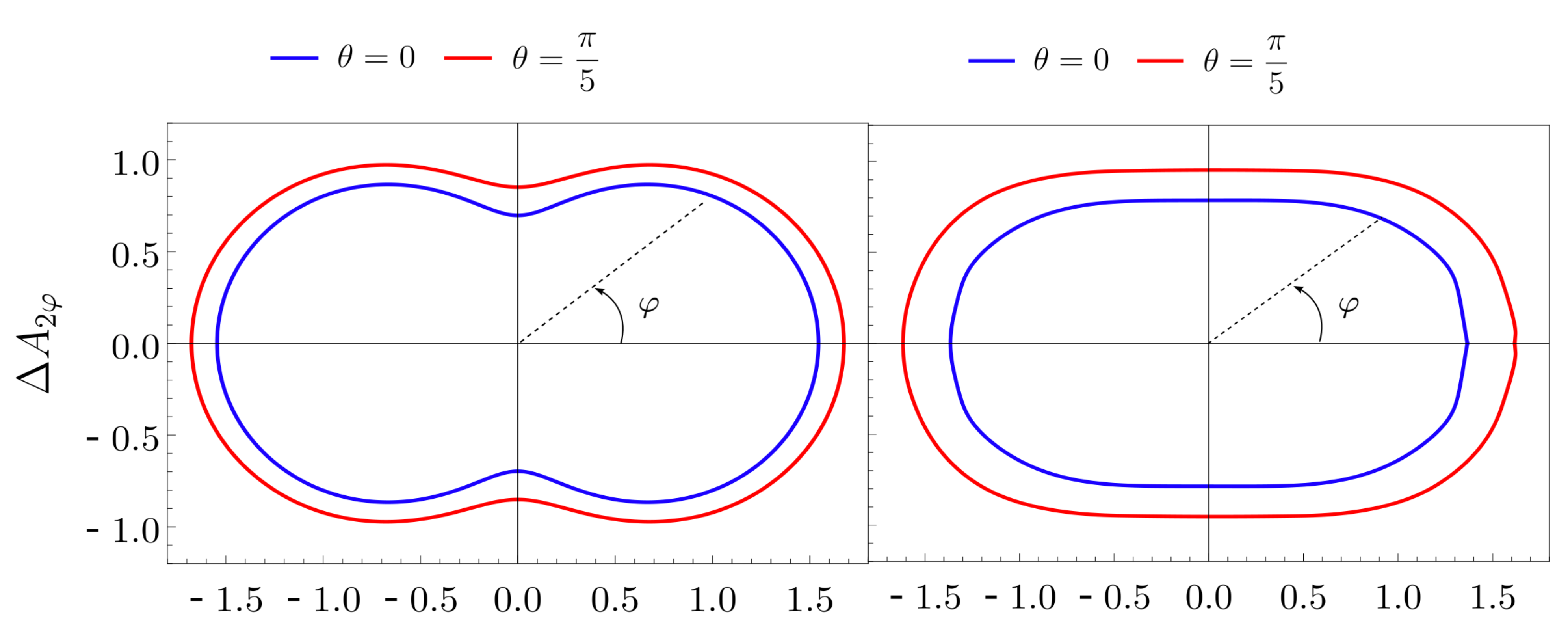}
\caption{(Color on-line) Left panel: polar plot of the quadratures fluctuations $\Delta A_{2\varphi}$, in unit of $\mathcal{A}(\omega)$ and for $q=1$, with respect to the phase $\varphi$, according to Eq.~\eqref{phi}. The injection of particles is done by means of a Lorentzian voltage drive with $\eta=0.1$. The blue line describes the non interacting case ($\theta=0$) while the red one stands for $\theta=\pi/5$. Right panel: polar plot of the quadratures fluctuations $\Delta A_{2\varphi}$ in the case of injection of particles through cosine voltage drives, in unit of $\mathcal{A}(\omega)$ and for $q=1$, with respect to the phase $\varphi$. The blue line describes the non interacting case ($\theta=0$) while the red one stands for $\theta=\pi/5$. Other relevant parameters are: $L=2.5\, \mu\mathrm{m}$, $v=2.8\times 10^4\, \mathrm{m/s}$, $\alpha=2.1$ and $\lambda=2$.} 
\label{Fig6}
\end{figure} 
In Figure~\ref{Fig6} it is shown the evolution of $\Delta A_{2\varphi}$, for $q=1$, when the phase $\varphi$ is varied in a polar plot. In the left panel we consider the periodic injection of Levitons in the non-interacting case (blue) and for $\theta=\pi/5$ (red) while in the right one we consider a cosine drive for $\theta=0$ (blue) and $\theta=\pi/5$ (red). From both panels, when we consider a fixed $\varphi$, we obtain the values for the fluctuations of the two orthogonal quadratures $\Delta A_{2\varphi}$ and $A_{2(\varphi+\frac{\pi}{2})}$ of the emitted electromagnetic field. In the left panel (Lorentzian voltage pulses), for the specific case of $\varphi=0$ and for $q=1$ we immediately observe that the quantum fluctuation $\Delta A_{\pi}$ reaches its minimum value at the expense of $\Delta A_0$. Thanks to the particular "peanut shape" of the left panel's Figure~\ref{Fig6}, it is clear that the maximum squeezing effect is obtained for the previous considered condition where $\Delta A_{\pi} $ is well below the value of quantum vacuum at the expenses of $\Delta A_0 $. This happens in order to preserve the Heisenberg's uncertainty relation in Equation~\eqref{heisen}. The same considerations about the shape of the curves can be done for both the non-interacting and interacting cases. The latter case leads to greater values of the quadratures' fluctuations but the squeezing property is still present even in the presence of interactions. This is in perfect agreement with what observed in Figure~\ref{Fig4} where even if $\langle A^2_\pi \rangle$, for $\theta=\pi/5$ and $q=1$, is bigger than $\langle A^2_\pi \rangle$ (for $\theta=0$ and $q=1$) the squeezing effects still remains. The right panel of Figure~\ref{Fig6} shows the polar plot of $\Delta A_{2\varphi} $ for a cosine drive, considered in Equation~\eqref{cosine}. In this panel, the shape of both non-interacting (blue) and interacting (red) cases is no longer similar to the "peanut" one. We can observe that for non-interacting channels the maximum squeezing is achieved for a fixed $\varphi=0$ while in the interacting case this quantum property is quite totally suppressed. According to this, we can remark that the Lorentzian shape of injection voltage is a better drive, even in the presence of interactions, in order to obtain a squeezed state of the emitted microwave radiation. Until now, we have assumed $\eta=0.1$ for the Lorentzian case.\\
\begin{figure}[ht]
\centering
\includegraphics[scale=0.4]{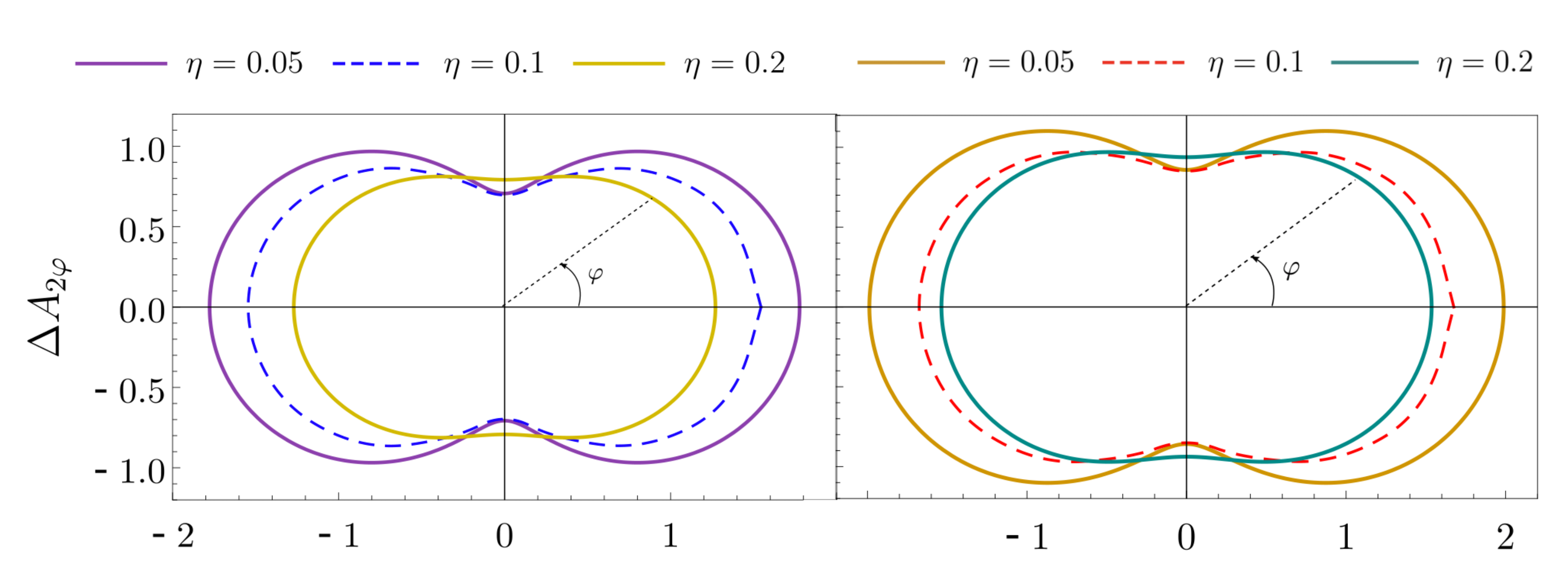}
\caption{(Color on-line) Polar plot of the quadratures fluctuations $\Delta A_{2\varphi}$ with respect to the angle $\varphi$, in unit of $\mathcal{A}(\omega)$. The voltage drive considered for the injection is a train of Lorentzian pulses with $q=1$. In the left we consider the non interacting case $\theta=0$ while in the right one we consider the interaction parameter $\theta=\pi/5$. The three different curves (in both panels) stand for three values of the ratio $\eta$: the purple (left) and gold (right) lines describes the case at $\eta=0.05$, the blue (left) and red (right) dashed lines are obtained for $\eta=0.1$ and the yellow (left) and light blue (right) represent the fluctuations for $\eta=0.2$. Other relevant parameters are: $L=2.5 \mu\mathrm{m}$, $v=2.8\times 10^4\, \mathrm{m/s}$, $\alpha=2.1$ and $\lambda=2$.} 
\label{Fig7}
\end{figure} 
Now, in Figure~\ref{Fig7} we compare the polar plot of quantum  fluctuations $\Delta A_{2\varphi}$ for a Lorentzian drive ($q=1$) with different values of the ratio $\eta$ between the width of the Lorentzian $w$ and the period $\mathcal{T}$. In the left panel we show the fluctuations in the non-interacting case while in the right one we consider the interaction $\theta=\pi/5$ between edge channels. If we consider an experimentally more challenging situation where the Lorentzian pulses are very narrow (purple curve with $\eta=0.05$), we get a more pronounced "peanut shape" (violet and gold line in both panels), but the squeezing effect doesn't improve with respect to the case $\eta=0.1$. Conversely, by increasing the ratio (yellow and light blue curves with $\eta=0.2$), the "peanut shape" disappears. In this last situation the effects of the squeezing is evidently reduced and tends to disappear in the interacting case with respect to the non-interacting one, similarly to what happens for a cosine drive.
\section{Conclusions}\label{conclusions}
We have theoretically investigated the dynamical response of the outgoing current fluctuations, which are associated to the photo-assisted finite frequency noise, in a QPC geometry for a QH bar at filling factor $\nu=2$ in presence of electron-electron interactions between the two channels. We have considered the periodic injection of excitations (in the GHz range) in the mesoscopic system by means of two different time dependent voltage drives, respectively a cosine and a train of Lorentzian pulses, and a short-range coupling between edge channels. We have shown that the effects of interaction lead to an unavoidable reduction of the squeezing of the electromagnetic quadratures for both drives. However, the Lorentzian case still remain a better candidate in order to achieve a relevant squeezing also in presence of interaction. We have investigated the quantum properties of the emitted light by studying how the quantum fluctuations vary in terms of the experimental tunable phase $\varphi$. Furthermore, we have shown the connection between the shape of the squeezed quantum state and the different regimes in the two filters set-up accessible in experiments. A deep knowledge of quantum properties of light is essential in the development of quantum information protocols and our work aims at contributing to this field by studying the role played by interactions on the emitted quantum radiation.
\section*{Acknowledgement} Authors would like to thank H. Bartolomei and G. F\`eve for useful discussions. This paper has been published thank to the funding of "Dipartimento di Eccellenza MIUR 2018-2022". 
\appendix
\section{Calculation of the dynamical response of the noise}
\label{appendixA}
In this Appendix we evaluate explicitly the expression of the dynamical response of the noise $\mathcal{X}_\varphi^{(k)}(\omega_0,\omega)$ appearing in Equation~\eqref{Xp_extended} in the zero temperature limit, according to what happens in the experimental situations. In order to reach this goal we have to calculate the correlator in Equation~\eqref{corr}
\begin{equation}
    \begin{split}
    \mathcal{X}_{+,\varphi}^{(k)}(\omega_0,\omega)&=e^{i\varphi}\langle I_1(\omega)I_1(k\omega_0-\omega) \rangle_c= \\
    &=e^{i\varphi}\int_{-\infty}^{+\infty} dt\, e^{i\omega t} \int_{-\infty}^{+\infty} dt'\,e^{i(k\omega_0-\omega)t'} \langle I_1(t)I_1(t') \rangle_c.
    \end{split}
\end{equation}
The current operator on the inner channel reads $I_1(t)=-ev_1 :\psi_1^\dagger(t)\psi_1(t):$, where $:...:$ denotes the normal ordering with respect to the Fermi sea and the fermionic fields operators $\psi_{1}$ are evaluated at the level of the two-filters set-up after the QPC.\\
By considering the following change of variables $\bar{t}=\frac{t+t'}{2}$ and $\tau=t-t'$, the previous equation can be rewritten as
\begin{equation}
    \mathcal{X}_{+,\varphi}^{(k)}(\omega_0,\omega)=e^{i\varphi}\int_{-\frac{T}{2}}^{+\frac{T}{2}} \frac{d\bar{t}}{T} \int_{-\infty}^{+\infty} d\tau e^{i\omega(\bar{t}+\frac{\tau}{2})}e^{i(k\omega_0-\omega)(\bar{t}-\frac{\tau}{2})} \langle I_1(\bar{t}+\frac{\tau}{2})I_1(\bar{t}-\frac{\tau}{2})\rangle_c \,.
    \label{a2}
\end{equation}
The effect of the external voltage drive $V_{\mathrm{in}}(t)$ can be properly taken into account with a phase factor in such a way that the correlation function, for the upper inner channel in Figure~\ref{Fig1}, is written as~\cite{Grenier13}
\begin{equation}
\begin{split}
    G_1(\bar{t}+\frac{\tau}{2},\bar{t}-\frac{\tau}{2})&=\langle \psi_1^\dagger(\bar{t}+\frac{\tau}{2})\psi_1(\bar{t}-\frac{\tau}{2})\rangle_c=
    \\&=\mathcal{G}_{1,F}(\tau)(e^{-i\Phi_{1}(\bar{t}+\frac{\tau}{2},\bar{t}-\frac{\tau}{2})}-1)
    \label{newcorr1}
    \end{split}
\end{equation}
where $\mathcal{G}_{1,F}(\tau)\propto \frac{1}{\tau}e^{i\omega \tau}$ is the correlation function for the equilibrium state (i.e., when no drive is applied) and the phase factor is
\begin{equation}
    \Phi_1(t,t')=e\int_{t'}^{t} V_{1,\mathrm{out}}(\tau) d\tau
\end{equation}
where $V_{1,\mathrm{out}}$ takes the form in Equation~\eqref{Vout} and carry information about the effects of interactions. The fermionic operator $\psi_1$, outgoing from the QPC in the inner channel, is
\begin{equation}
    \psi_1(t)=\sqrt{\mathcal{T}}\,\psi_{1R}(t)+i\sqrt{\mathcal{R}}\,\psi_{1L}(t)
    \label{psiout}
\end{equation}
where $\psi_{1R}$ and $\psi_{1L}$ are the incoming inner ones. The previous relation descends directly from the scattering matrix in Equation~\eqref{qpc}.\\
In the Hanbury-Brown-Twiss (HBT) geometry considered in this manuscript and depicted in Figure~\ref{Fig1}, after substituting~\eqref{psiout} in Equation~\eqref{newcorr1} the HBT contribution to the correlation function is 
\begin{equation}
    G_1(\bar{t}+\frac{\tau}{2},\bar{t}-\frac{\tau}{2})=-i\sqrt{\mathcal{R}\mathcal{T}}\mathcal{G}^{(R,L)}_{1,F}(e^{-i\Phi_1(\bar{t}+\frac{\tau}{2},\bar{t}-\frac{\tau}{2})}-1).
    \label{a6}
\end{equation}
By substituting Equation~\eqref{a6} in~\eqref{a2}, the current correlator becomes 
\begin{equation}
    \langle I_1(\bar{t}+\frac{\tau}{2})I_1(\bar{t}-\frac{\tau}{2})\rangle_c=(e v_1)^2\, \mathcal{R}\mathcal{T}\, \mathcal{G}_{1,F}^{(R,L)}(\tau)\mathcal{G}_{1,F}^{(R,L)}(\tau)\,\left[e^{-i \Phi_{1}(\bar{t}+\frac{\tau}{2},\bar{t}-\frac{\tau}{2})}+e^{+i \Phi_{1}(\bar{t}+\frac{\tau}{2},\bar{t}-\frac{\tau}{2})}-2\right]
\end{equation}
where $\mathcal{R}$ and $\mathcal{T}$ stand for the reflection and transmission probability of the QPC and $(R,L)$ indicate the right- and left-movers along the inner channels of the two edges. For a periodic voltage pulse source, considered here, the phase factor can be written in terms of the Fourier series
\begin{equation}
    e^{-ie\int_{0}^{t}V_{1,\mathrm{out}}(\tau)d\tau}=e^{-i q\omega t}\sum_{l} \widetilde{p}_{l}(z)e^{-i l\omega_0 t}
    \label{fourier}
\end{equation}
where $\omega_0=2\pi/T$ (with $T$ period of the voltage source), $z=e\tilde{V}/\omega_0$ and $q=eV_{\mathrm{DC}}/\omega$. The explicit expressions of the coefficients $\widetilde{p}_{l}(z)$ is
\begin{equation}
    \widetilde{p}_{l}(z)=\sum_{n}p_{l-n}(z_1)p_n(z_2)e^{i\omega_0\tau_\rho(l-n)}e^{i\omega_0 \tau_\sigma (n)}
\end{equation}
which is calculated in Reference~\cite{Rebora20} and where $z_1=z \cos^2\theta$ and $z_2=z \sin^2\theta$ take into account the effects of interactions through $\theta$. Furthermore the form of the coefficients $p_l$ is related to particular form of the drive considered (cosine or Lorentzian) and it is expressed in Equations~\eqref{cos} and~\eqref{lor}. By substituting Equation~\eqref{fourier} in~\eqref{a2} we have
\begin{equation}
    \begin{split}
        \mathcal{X}^{(k)}_{+,\varphi}(\omega_0,\omega)=&(ev_F)^2\mathcal{R}\mathcal{T}e^{i\varphi}\int_{-\frac{T}{2}}^{+\frac{T}{2}} \frac{d\bar{t}}{T}\int_{-\infty}^{+\infty} d\tau
\frac{1}{\tau^2} e^{i(\omega-\frac{k\omega_0}{2})\tau}e^{ik\omega_0 \bar{t}} \times \\
&\times \sum_{l l'} \widetilde{p}_{l}(z) \widetilde{p}^*_{l'}(z)e^{-i\omega_0(l-l')\bar{t}} e^{-i\omega_0(l+l')\frac{\tau}{2}}e^{i q\omega \tau}.
\end{split}
\end{equation}
After the first integration in $\bar{t}$ the previous Equation gives a $\delta(l-l'-k)$ and the expression becomes
\begin{equation}
    \begin{split}
        \mathcal{X}^{(k)}_{+,\varphi}(\omega_0,\omega)&=(ev_F)^2\mathcal{R}\mathcal{T}e^{i\varphi}\int_{-\infty}^{+\infty} d\tau \frac{1}{\tau^2}\sum_{l} \widetilde{p}_{l}(z) \widetilde{p}^*_{l-k}(z) e^{-i\omega_0\frac{\tau}{2}(2l-k)}e^{iq\omega \tau} e^{-i k\omega_0 \tau/2} e^{i\omega\tau}\\
        &=(ev_F)^2\mathcal{R}\mathcal{T}e^{i\varphi}\sum_l(\pi|\omega_0 l+q\omega-\omega|)\\
        &=(ev_F)^2\mathcal{R}\mathcal{T}e^{i\varphi}\sum_l(\pi\omega|\lambda l+q-1|)
        \end{split}
\end{equation}
where we have used the short hand notation $\lambda=\omega_0/\omega$ and $|...|$ denotes the absolute value. The same steps can be followed in order to calculate the other term $\mathcal{X}_{+,\varphi}^{(-k)}(\omega_0,-\omega)$ in Equation~\eqref{Xp}. The above (infinite) sum is convergent and its value has been obtained numerically by summing over a finite number of coefficients until the desired precision is obtained.
\section{Comparison of the Lorentzian and cosine drive at the same $z$}
\label{app2}
\begin{figure}[ht]
\centering
\includegraphics[scale=0.37]{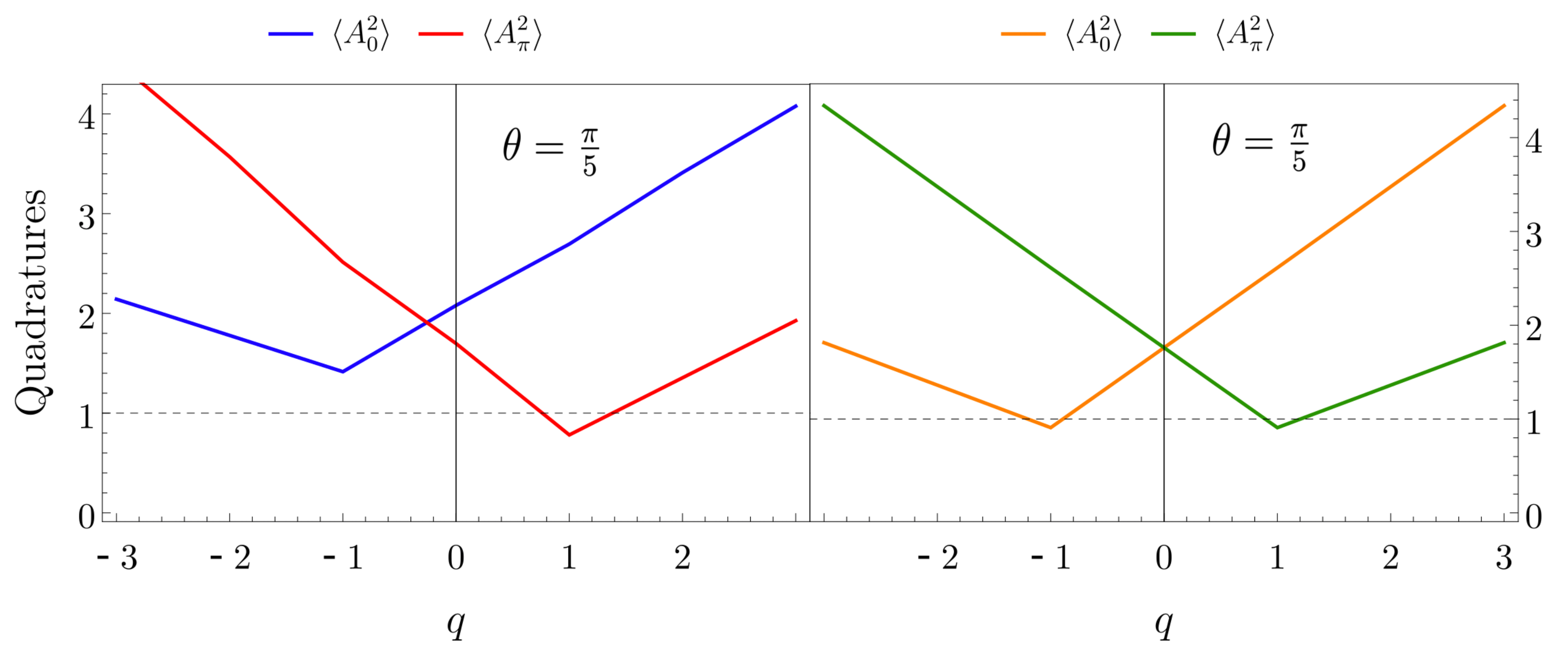}
\caption{(Color on-line) Quadratures of the emitted electromagnetic field in units of $\mathcal{A}(\omega)$ as a function of the number of injected electrons $q=eV_{\mathrm{DC}}/\hbar\omega$ for a Lorentzian drive (left panel) and a cosine drive (right panel). For both panels we have considered the same interaction strength ($\theta=\pi/5$) and the same $z=e\tilde{V}/\hbar\omega_0=1.32$. This value is the one which minimize the noise for the cosine drive (see Figure~\ref{Fig3} in the main text). However, the squeezing associated to the Lorentzian drive is still slightly better than the cosine one for $\langle A^2_\pi \rangle$ quadrature. The dashed gray horizontal line indicates the vacuum fluctuations. For the Lorenztian drive we have chosen $\eta=0.1$. Other parameters are: $L=2.5\,\mu\mathrm{m}$, $v=2.8\times 10^4\, \mathrm{m/s}$, $\alpha=2.1$ and $\lambda=2$.} 
\label{Fig8}
\end{figure}

The results presented in Section~\ref{results} have been obtained by a numerical minimization of the noise as a function of $z=e\tilde{V}/\hbar\omega_0$. This procedure is done in order to compensate the effects of interactions, which deform the pulse of the AC drive, and to find the optimal conditions leading to the maximal squeezing. In this Appendix we compare the quadratures of the emitted electromagnetic field for a Lorentzian and a cosine drive at the same value of $z$. In Figure~\ref{Fig8} we present the two orthogonal quadratures $\langle A^2_0 \rangle$ and $\langle A^2_\pi \rangle$ of the emitted electromagnetic field for the Lorentzian drive (left panel) and the cosine one (right panel) at the same interaction strength $\theta=\pi/5$ and at the same $z=1.32$. This particular value of $z$ is the one which minimize the noise for the interacting cosine case (see Figure~\ref{Fig3}). It is worth noting that even considering the optimal value of $z$ for the cosine pulse, the squeezing associated to the Lorentzian one is still slightly better.

\end{document}